\documentclass[aps,nofootinbib,notitlepage,longbibliography,twocolumn, superscriptaddress]{revtex4-1}
\usepackage{amsmath,amssymb}
\usepackage{slashed}
\usepackage{graphicx,xcolor}
\usepackage{bm}
\usepackage[T1]{fontenc}
\usepackage[utf8]{inputenc}
\usepackage{gauss} 
\usepackage[top=1.0in,bottom=1.0in,left=1.0in,right=1.0in]{geometry}
\usepackage[colorlinks,linkcolor=blue,citecolor=blue]{hyperref}
\usepackage{subfig}
\newcommand{\be}{\begin{equation}}
\newcommand{\ee}{\end{equation}}
\newcommand{\bea}{\begin{eqnarray}}
\newcommand{\eea}{\end{eqnarray}}
\newcommand{\ba}{\begin{eqnarray}}
\newcommand{\ea}{\end{eqnarray}}
\DeclareMathOperator\arctanh{arctanh}

\def\be{\begin{eqnarray}}
\def\ee{\end{eqnarray}}
\def\bea{\be}
\def\eea{\ee}

\def\roughly#1{\mathrel{\raise.3ex\hbox{$#1$\kern-.75em%
\lower1ex\hbox{$\sim$}}}}

\usepackage{lipsum}

\newcommand{\dd}{\mathrm{d}}

\date{\today}
\begin{abstract}
 We revisit the entanglement of a Schwinger pair created by external fields of arbitrary 
strength, using a holographic dual description of QCD. When external fields are strong in comparison
to the string tension, the entanglement is  geometrically tied  to the Einstein-Rosen
(ER) bridge in the bulk, and disappears when the pair production is not exponentially suppressed at the boundary.  
For moderate external fields,  the entanglement is shown to follow from the geometrical interplay between  the position of the ER bridge and the confining wall in the bulk. We clarify the physical  nature of quantum entanglement in pair production, and connect it to the entropy of entanglement between the left- and right-moving fermions. In particular, we clarify the effect of real radiation off the produced particles on quantum entanglement of the pair.
\end{abstract}
\begin{document}
\title{Entanglement in a holographic Schwinger pair with confinement}
\author{Sebastian Grieninger}
\email{sebastian.grieninger@stonybrook.edu}
\affiliation{Center for Nuclear Theory, Department of Physics and Astronomy,
Stony Brook University, Stony Brook, New York 11794–3800, USA}

\author{Dmitri E. Kharzeev}
\email{dmitri.kharzeev@stonybrook.edu}
\affiliation{Center for Nuclear Theory, Department of Physics and Astronomy,
Stony Brook University, Stony Brook, New York 11794–3800, USA}
\affiliation{Department of Physics, Brookhaven National Laboratory
Upton, New York 11973-5000, USA}

\author{Ismail Zahed}
\email{ismail.zahed@stonybrook.edu}
\affiliation{Center for Nuclear Theory, Department of Physics and Astronomy,
Stony Brook University, Stony Brook, New York 11794–3800, USA}
\maketitle

\section{Introduction}
Entanglement 
underlies the quantum description of physical laws. 
Quantum states are complex superposition states in a Hilbert space, with a 
content that is revealed by the projection measurement. As a result, 
two non-causally related measurements can be correlated.

A striking example of these non-causal correlations is the famous Einstein-Rosen-Podolsky (EPR) paradox. 
The common realization of this paradox is that of an entangled pair of spin-$\frac 12$ in an initial 
spin singlet state, with spins that are correlated at space-like separations.
In particular, this canonical setup allows to derive the simplest Bell inequality that 
 rules out the concept of hidden variables in quantum mechanics.

Recently, the quantum entanglement in Schwinger pair creation process~\cite{Schwinger:1951nm}  was described using a gravity dual description~\cite{Jensen:2013ora,Sonner:2013mba}. 
Remarkably, the dual of the EPR pair in bulk is a string
~\cite{Xiao:2008nr,Semenoff:2011ng,Lewkowycz:2013laa,Chernicoff:2013iga,Jensen:2014bpa,Hubeny:2014zna,Ghodrati:2015rta,Semenoff:2018ffq} 
with a worldsheet that sustains a non-traversable wormhole (Einstein-Rosen (ER) bridge), suggesting the equivalence between the entanglement and the wormholes within the holographic correspondence~\cite{Maldacena:2013xja}.
The extension to traversable wormholes was further worked out in~\cite{deBoer:2022rbn}. The possible applications entanglement in pair production to  phenomenology of high energy hadron and nuclear collisions were discussed in~\cite{Kharzeev:2005iz,Muller:2022htn}.

In the Schwinger quark-antiquark pair production, the end-points are never causally connected.
Nevertheless, the pair is entangled due to its overall color neutrality. In non-confining dual gravity
description, the entanglement was found to be of order $\sqrt{\lambda}$ in the weak field limit~\cite{Jensen:2013ora,Hubeny:2014zna,Hubeny:2014zna}, where $\lambda=g_Y^2N_c$ is the $'$t Hooft strong coupling. 

Outside of the holographic framework, the entanglement in Schwinger pair production was addressed in \cite{Florio:2021xvj}, without considering the radiation emitted by the produced particles. It has been found there that the entropy of entanglement between the left-moving antifermions and the right-moving fermions is exactly equal to the statistical Gibbs entropy of the produced state. It is of great interest to check whether this intriguing relation is modified in the presence of radiation off the produced fermions. Recent quantum simulations indicate that radiation off the produced pair reduces the amount of entanglement \cite{Florio:2023dke}.

In this work, we will extend the holographic treatment of Schwinger pair production to the external U(1) fields of arbitrary strength, and consider also the effects of confinement. The first study of the holographic Schwinger process in a confining background was carried out in~\cite{Kim:2008zn}, where a critical threshold for the electric field was noted. The case of pair production in a confining geometry was studied also in ~\cite{Liu:2018gae}, and it was found that the wormhole does not appear when the quark and antiquark are separated in the transverse space. The emergence of horizon, and the ensuing interpretation of Schwinger pair production in terms of thermal Hawking-Unruh radiation was discussed in \cite{Kharzeev:2005iz,Volovik:2022cqk}.

The organization of the paper is as follows: In section~\ref{SEC_0} we briefly outline the effect of confinement on the Schwinger pair creation process. Confinement inhibits the pair production, unless the applied electric field is sufficiently strong to overcome the effect of the confining string. 
In section~\ref{SEC_II} we discuss the holographic setup in the weak field limit
for Schwinger pair production.  Mikhailov's light-like surface~\cite{Mikhailov:2003er} for a quark and anti-quark receding
at a constant light-like acceleration in conformal AdS$_5$  is used to evaluate the bulk action explicitly. The emerging Unruh 
temperature allows for the extraction of the free energy. We identify the quantum entanglement entropy with the ensuing thermal
entropy. The results are readily extended to ``walled" AdS$_5$ to account for confinement.  In section~\ref{SEC_III} we extend the 
holographic setup to the strong field limit, and derive the quantum entanglement entropy for the conformal and confining geometries
in curved AdS. In section~\ref{LUNDX} we make a link to phenomenology by using a schematic form of the Lund model to derive the entanglement entropy associated to the string fragmentation in jets. This entanglement entropy measures the Schwinger pair creation probability. This entropy is smaller than the single tunneling entanglement entropy from the confining geometry, owing to the decoherence due to the random tunnelings, yet it should be accessible to experimental measurements. Our conclusions are in section~\ref{SEC_IV}. 

\begin{figure}
    \centering
    \includegraphics[width=0.6\linewidth]{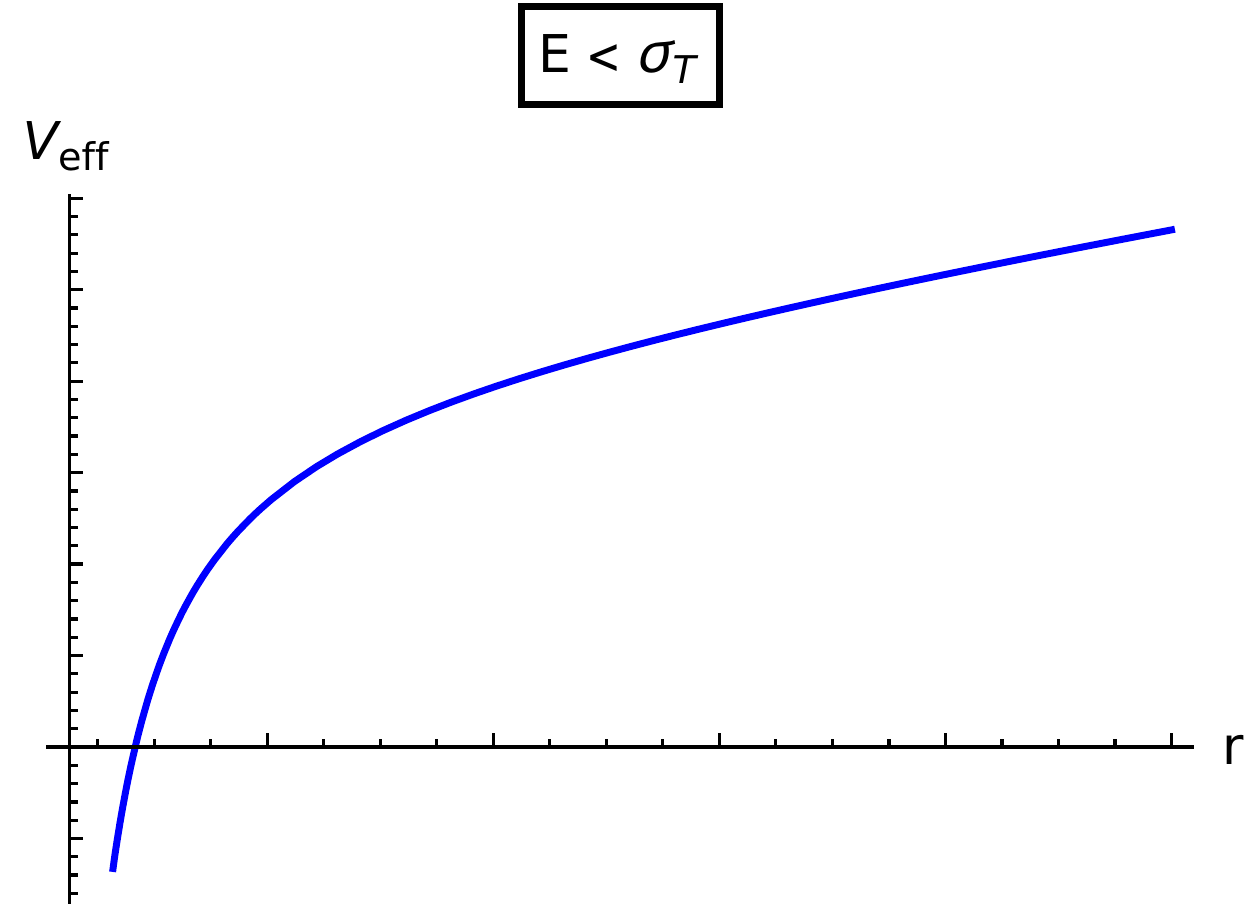} \\ 
    \includegraphics[width=0.6\linewidth]{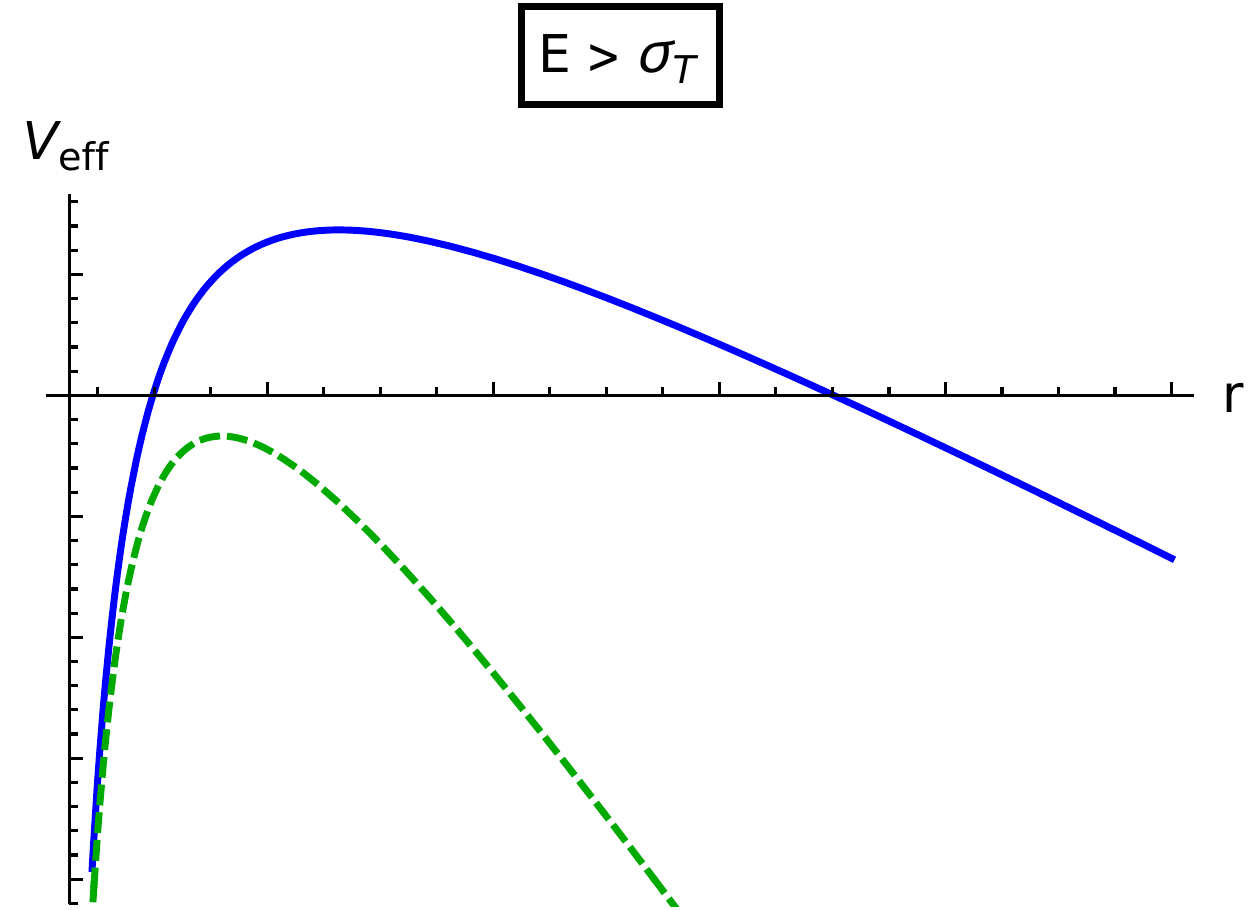}
    \caption{Effective potential $V_\text{eff}(r)$ versus the distance, for weak  (top) and strong (bottom) electric field ($E_c>E>\sigma_T$: blue, $E>E_c>\sigma_T$: green, dashed).}
    \label{fig:Veff}
\end{figure}

\section{Pair creation}
\label{SEC_0}
Before we detail the holographic analysis of the Schwinger pair creation process in a confining AdS background, let us discuss the general features of this process. It is clear that to produce a pair of quarks with mass $M$ from the vacuum, 
an electric field $E$ should lose energy in excess of  $2M$. In the presence of confinement described by a linear quark-antiquark potential $\sigma_T r$ (where $\sigma_T$ is the string tension), this requirement is made more prohibitive with the threshold set at $2M+\sigma_T \,r$,
with $r$ the pair separation. 
With this in mind, the effective potential for a 
$q\bar q$ pair undergoing pair production is typically of the form 
\bea
\label{X1}
V_{\rm eff}(r) =2M-(E-\sigma_T)\,r-\frac{C{\sqrt\lambda}}r ,
\eea
where we have added the attractive Coulomb-type quark-antiquark interaction that is important 
at short distances; $\lambda=g_Y^2N_c$ is the strong 't Hooft coupling, and the constant $C$ was fixed in~\cite{Maldacena:1998im}. Throughout this paper, we include the coupling $g_Y$ in the definition of $E$. 

The potential (\ref{X1}) is illustrated in Fig.~\ref{fig:Veff}. Clearly, for $E<\sigma_T$,
the confined pair cannot be separated and pair production is prohibited. Pair production is only possible if $E>\sigma_T$,
for which the tunneling through the potential barrier is exponentially suppressed. For sufficiently strong fields,
$E>E_c>\sigma_T$, the barrier vanishes, and pair production 
is no longer penalized. Without confinement (when $\sigma_T=0$), a simple  estimate of $E_c$ can be made by assuming that the work done by electric field $E_c r$ should match $2M$ at distance where the energy of the strong Coulomb attraction equals $2M$, 
\bea
E_c r \sim 2M \sim \frac{C\sqrt\lambda}{r}. 
\eea 
This relation yields $r \sim C\sqrt\lambda /2M$, and 
\bea
E_c\sim  \frac{4M^2}{C\sqrt\lambda} .
\eea
With confinement, the critical value 
is shifted by the string tension, $E_c\rightarrow E_c+\sigma_T$.

\section{Holographic setup: weak field}
\label{SEC_II}
To address the pair production in a holographic setup modeling QCD, we will use a bottom-up approach. In the double limit of large $N_c$ (at fixed number of flavors $N_f$) and strong 't Hooft coupling $\lambda$, 
quarks are permanently confined with no string breaking. Large Wilson loops exhibit an area law at all distances. In the gravity dual, the area law follows from a simple AdS$_5$ with a hard wall \cite{Erlich:2005qh},  with a line element
\bea
\label{ADSWALL}
    \dd s^2=\frac{L^2}{z^2}((\dd x^\mu)^2+\dd z^2)\qquad z\leq z_H.
\eea
and zero otherwise. The string tension is readily found to be
\bea
\label{SIGMAT}
\sigma_T=\frac 1{2\pi \alpha^\prime}\frac{L^2}{z_H^2}=
\frac{\sqrt\lambda}{2\pi} \frac 1{z_H^2}.
\eea
Here $\alpha^\prime$ is the Regge slope for an open string, fixed by the $\rho$ meson trajectory $\alpha^\prime= 1/2m_\rho^2$. Inside the QCD string,  the strength of the chromoelectric field is proportional to the string  tension  $E\simeq 4\sigma_T$~\cite{Casher:1978wy}.

For  $z_H\rightarrow \infty$ at fixed coupling, the string tension vanishes and (\ref{ADSWALL}) reduces to conformal AdS$_5$. 
We will start our discussion with this latter case to clarify some issues regarding the evaluation of the entanglement entropy, and then proceed to the pair creation process in the confining case. 

\begin{figure}
    \centering
    \includegraphics[width=0.6\linewidth]{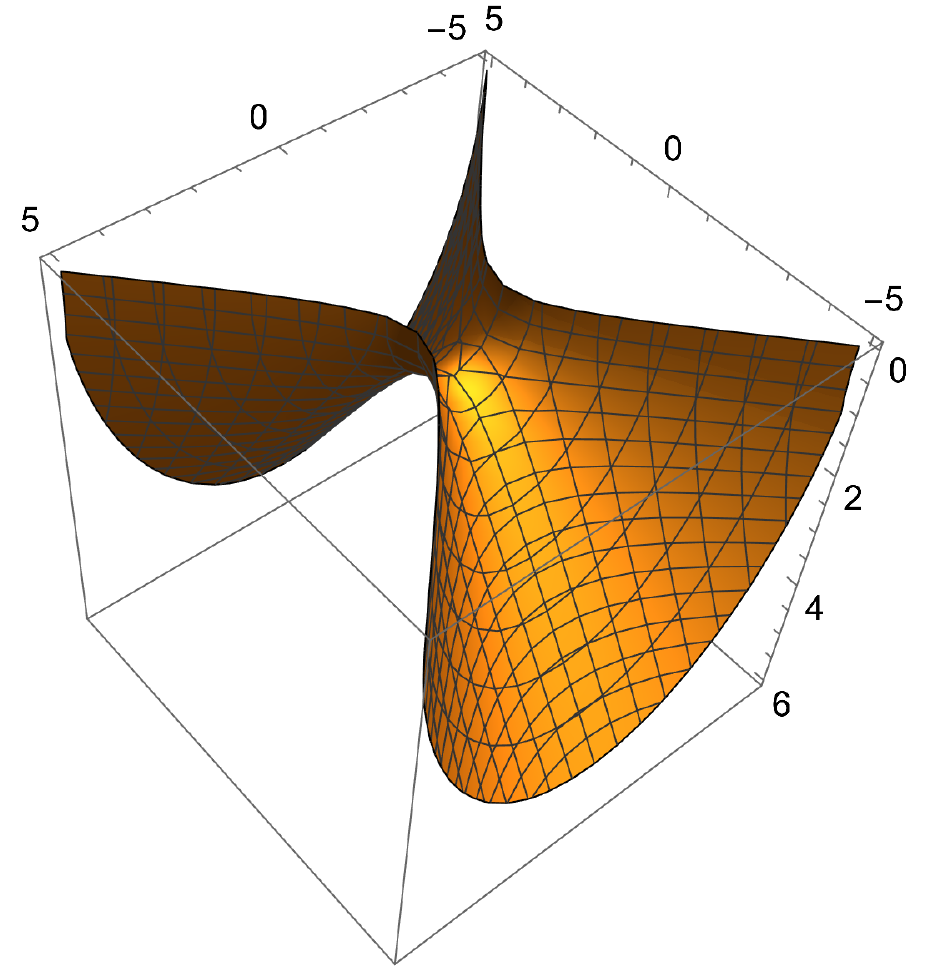}
    \caption{Hyperboloid worldsheet $-t^2+x^2+z^2=\frac 1{a^2}\equiv 1$ for $|x|,|t|\leq 5$ and $0\leq z\leq 6$.}
    \label{fig:EE3D}
\end{figure}

\subsection{AdS$_5$}
The holographic Schwinger effect in conformal AdS was discussed in~\cite{Semenoff:2011ng,Semenoff:2018ffq},
using a string worldsheet in Euclidean signature. In
Minkowski signature, the worldsheet is the ruled surface of delayed rays in the bulk~\cite{Mikhailov:2003er,Xiao:2008nr,Ghodrati:2015rta,Garcia:2012gw}, 
\begin{equation}
\label{XMTZ}
    X^M(\tau,z)=
    (z\dot x^\mu(\tau)+x^\mu(\tau),z)
\end{equation}
with the 4d  hyperbolic trajectories on the boundary
\bea
\label{BOUNDX}
x_\pm^\mu(\tau)=\bigg(\frac 1a{\rm sinh}(a\tau),\pm \frac 1a {\rm cosh}(a\tau), 0,0\bigg)
\eea
for constant acceleration $a=E/M$. The worldsheet (\ref{XMTZ}-\ref{BOUNDX}) is a hyperboloid
\bea
\label{XHYPER}
X_M^2=-t^2+x^2+z^2=x_\mu^2(\tau)=\frac 1{a^2}
\eea
as illustrated  in Fig.~\ref{fig:EE3D} for  $a^2=\ddot x^2\equiv1$. 
The  line element associated to (\ref{XMTZ}) is
\begin{equation}
    \dd X^M=(\dd z\,\dot x+z\,a\,\dd\tau+\dot x\,\dd\tau,\dd z).
\end{equation}
The determinant of the induced metric of the worldsheet embedded in AdS$_5$ is given by
\begin{widetext}
\begin{equation}
\label{ZDETX}
  \det\left( g_{mn}\frac{\partial X^m}{\partial \nu}\frac{\partial X^n}{\partial \sigma}\right)=\det\left(\begin{matrix}\frac{L^2}{z^2}(za+\dot x)^2&\frac{L^2}{z^2}\dot x(za+\dot x)\\
 \frac{L^2}{z^2} \dot x(za+\dot x)& \frac{L^2}{z^2}(\dot x^2+1)
   \end{matrix}\right)=\det\left(\begin{matrix}L^2(a^2-\frac{1}{z^2})&-\frac{L^2}{z^2}\\
 -\frac{L^2}{z^2} & 0
   \end{matrix}\right),
\end{equation}
\end{widetext}
since $\dot x^2=-1$ and $0=\dd/\dd\tau (\dot x^2)=2a\dot x$. 

Remarkably, (\ref{XMTZ}) can be regarded
as the locus of the radiated gluons between the receding pairs at strong coupling, as captured by dual gravity, with the power radiated at the boundary~\cite{Mikhailov:2003er,Xiao:2008nr}
\bea
\label{NRAD}
\frac{d{\cal E}}{dt}=\frac{\sqrt\lambda}{2\pi} a^2
\eea
This is to be compared to the weak coupling, classical dipole-like Larmor emission with the total radiated power $P$  
\bea
\label{LARMOR}
P=\frac 23 g_Y^2  a^2
\eea
\\
\\
{\bf Causal contribution:}
\\
From (\ref{ZDETX}) we note the appearance of a horizon at $z=\frac 1a$, whereby the
produced pair loses causal contact~\cite{Xiao:2008nr,Jensen:2013ora}. With this in mind,
the Nambu-Goto action  supported by the causal part of the worldsheet is
\bea
\label{DS}
    &&\Delta S=2\times\frac{1}{2\pi\alpha^\prime}\int\limits_{-\mathcal T/2}^{\mathcal T/2}\dd\tau\,\int\limits_{z_M}^{1/a}\dd z\,\frac{\alpha'\,\sqrt{\lambda}}{z^2}\nonumber\\
    &&=\frac{\sqrt{\lambda}}{2\pi}\mathcal T\,2\left(\frac{1}{z_M}-a\right)=2\mathcal T\left(M-\frac{a\sqrt{\lambda}}{2\pi}\right),\nonumber\\\label{eq:eesqrtl}
\eea
after using (\ref{ZDETX}). 
\begin{figure}
    \centering
    \includegraphics[width=0.7\linewidth]{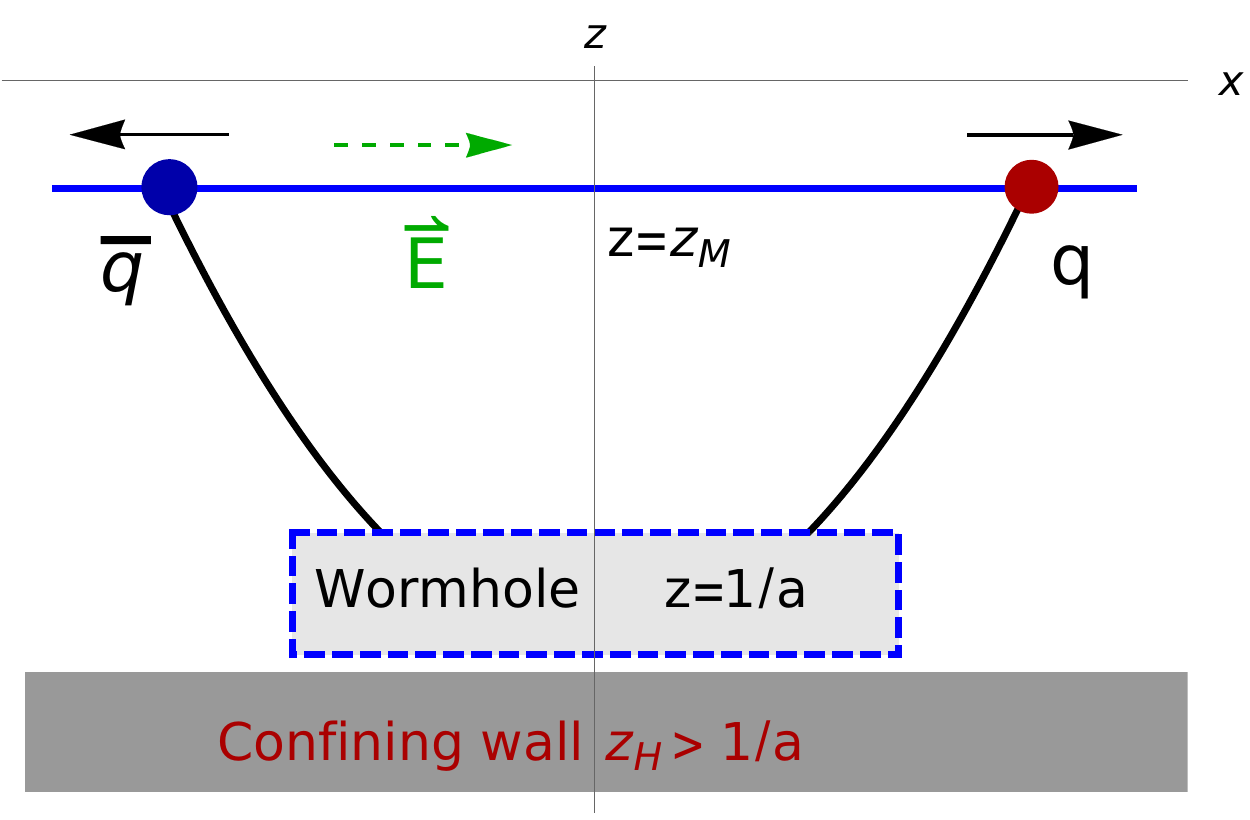} \\ 
    \includegraphics[width=0.7\linewidth]{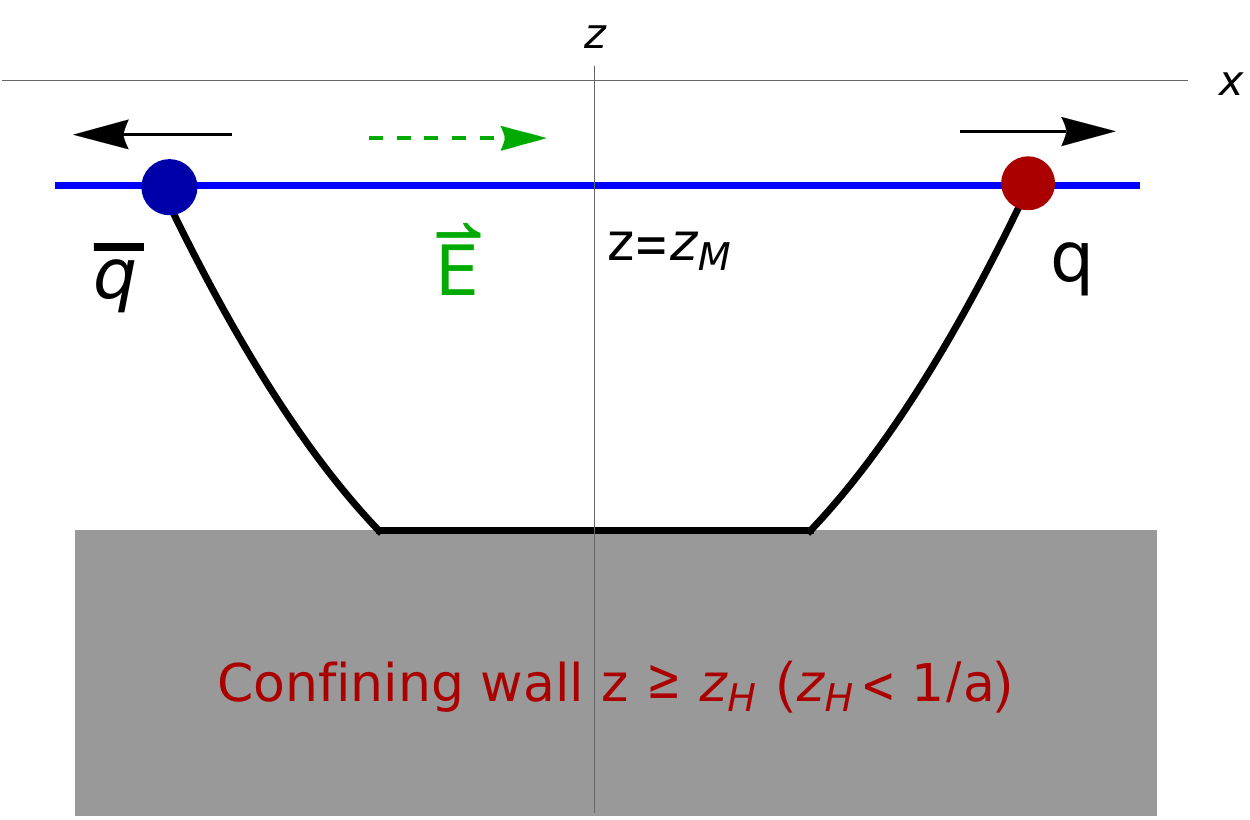}
    \caption{\textit{Top panel:} Non-confining case ($z_H>1/a$). The string world sheet exhibits a non-traversable wormhole with world sheet horizons at $z=1/a$. The hard wall does not affect the $q\bar q$ pair. \textit{Lower panel:} Confining case ($z_H<1/a$). The world sheet does not have a wormhole which is cut off by the hard wall at $z=z_H<1/a$.}
    \label{fig:ws}
\end{figure}
The overall factor of 2 counts the two separate sheet contributions in Fig.~\ref{fig:EE3D},
when cut at $\frac 1a$. The contribution from the boundary interaction term using (\ref{BOUNDX}),
is
\bea
\label{DSB}
&&\Delta S_B=\nonumber\\
&&-2\times \frac E2\int^{{\cal T}/2}_{-{\cal T}/2} d\tau (\dot{x}_+^0(\tau)x_+^1(\tau)-x_+^0(\tau)\dot{x}_+^1(\tau))\nonumber\\
&&\qquad\,\,\,=-M\cal{T}
\eea

Combining (\ref{DS}) with (\ref{DSB}) yields
the energy {\it{per particle}},
\bea
\label{ENERGY}
    \frac {\cal E}2=\left(M-\sqrt{\lambda}T_U -\frac M2\right),
\eea
with the Unruh temperature $T_U=a/(2\pi)$.
 The temperature correction
is reminiscent of the Debye correction to charge particle's self-energy
\bea
{\cal E}_D\sim -\lambda \frac{e^{-m_D r}}{r}\sim -\lambda m_D\sim -\lambda^{\frac 32} T_U\,,
\eea
at weak coupling, where we used $r\sim {m_D}^{-1}$, $m_D \sim \sqrt{\lambda}\ T_U$ for an estimate.
The entanglement entropy follows from (\ref{ENERGY}) if we interpret it as a free energy $F$ of the particle at Unruh temperature~\cite{Hubeny:2014zna,Semenoff:2018ffq}, and use thermodynamic relation $S_{EE} = - \partial F / \partial T$:
\begin{equation}
\label{XSEEX}
    S_\text{EE}=\sqrt{\lambda}.
\end{equation} 
\\
\\
{\bf Non-causal contribution:}
\\
The remainder of the action comes from the non-causal contribution and describes the radiation from the produced pair. We will now see that this radiation reduces the amount of entanglement. 

Let us begin by noting that the induced metric on the worldsheet
\begin{equation}
\label{DSIND}
    \dd X_M^2=L^2\left(a^2-\frac{1}{z^2}\right)\dd \tau^2-\frac{2\,L^2}{z^2}\dd\tau\,\dd z.
\end{equation}
may be written as the metric of a topological black hole~\cite{Emparan:1999gf,Casini:2011kv}. This is manifest, if we  set  $w=1/(az)$ and $\tau=t/a-1/a \arctanh{w}$~\cite{Hubeny:2014kma}, so that (\ref{DSIND}) reads
\begin{equation}
    \dd X_M^2=-(w^2-1)\,\dd t^2+\frac{\dd w^2}{w^2-1},
\end{equation}
which clearly exhibits a world-sheet horizon for $w=1$ (corresponding to $z=1/a$).
In Euclidean signature, we should require that the Wick rotated time (replica) coordinate is an angle with  $t=t+2\pi n$. We can achieve this by replacing the blackening factor $f(w)=-1+w^2$ by~\cite{Emparan:1999gf,Chalabi:2020tlw}
\begin{widetext}
\bea
    f_n(w)=w^2-1-\frac{(w_h^d-w_h^{d-2})}{w^{d-2}},\qquad\qquad w_h=\frac{\sqrt{1+n^2d(d-2)}+1}{nd},\label{eq:regbh}
\eea
\end{widetext}
where the horizon is located at $w=w_h$ and $d+1$ is the dimension of the gravity theory.

The regulated metric corresponds to the metric of a charged topological black hole, and the entropy is given by the volume of the horizon region~\cite{Casini:2011kv}. In fact, replacing the upper integration limit in~\eqref{eq:eesqrtl} by the regulated horizon~\eqref{eq:regbh}, taking the derivative with respect to $n$ and setting $n=1$ at the end, one gets~\cite{Chang:2013mca,Lewkowycz:2013laa,Karch:2014ufa,Chalabi:2020tlw,Ju:2023bjl}
\begin{equation}
\label{XSEE}
    S_\text{EE}=-\frac{\sqrt\lambda}{2\pi}\left(\partial_n(2\pi w_h)\right)_{|n=1}=
    \frac  {\sqrt\lambda}{d-1}.
\end{equation}
Alternatively, if we were to use the unregulated planar black hole (i.e. the hyperbolic black brane with the charge set to zero from the start) then we  have  $S_\text{EE}=\sqrt{\lambda}$. If we use the regulated metric, and send the charge to zero after taking the derivative with respect to the temperature, then  (\ref{XSEE}) follows.

If we recall the causal contribution (\ref{XSEEX}), then (\ref{XSEE}) in $d=4$ can be represented as
\bea
S_{EE}={\sqrt\lambda}-\frac 23 \sqrt\lambda,\label{eq:twocontr}
\eea
where the first contribution to the entanglement arises from the causal part of the
entangling surface (representing the dressing of the charge by virtual gluons), and the second contribution -- from the non-causal part of the entangling surface that represents the emitted radiation. In the holographic description, this radiation is hidden behind the worldsheet horizon (dual to the Rindler horizon in Minkowski space). We thus conclude that the emitted radiation reduces the amount of entanglement between the produced particles.

The reduction of the entanglement entropy in the presence of radiation can be expected 
if we correct the energy per particle (\ref{ENERGY}) for the energy radiated away. Using
(\ref{NRAD}), we find that the energy lost or radiated away from the charged particle  is
\bea
{\cal E}_{\rm loss}=\sqrt{\lambda}T_U \,a{\cal T}\sim \sqrt{\lambda}T_U ,
\eea
where we used for 
the duration of radiation ${\cal T}$ the time it takes to reach the worldsheet horizon,  ${\cal T}\sim \frac 1a$.

Our result~\eqref{eq:twocontr} clarifies the difference  between the result in~\cite{Jensen:2013ora}, which states that the entanglement entropy of the uniformly accelerated $q\bar q$ pair is $S_{EE}=\frac 13 \sqrt\lambda$, and the result in~\cite{Hubeny:2014zna} which finds $S_{EE}={\sqrt\lambda}$. The latter takes only the virtual  radiation  into account (shown in black in~\ref{fig:fey}), while the former  also accounts for the real radiation of the moving charged particles (shown in red in~\ref{fig:fey}), which causes energy loss and a decrease in  the net entanglement entropy.

\subsection{Walled AdS$_5$}
In order to have a confining background, we introduce a hard wall at $z=z_H$ as in (\ref{ADSWALL}). In case of $z_H\ge1/a$, our calculation reduces to the result of the last section as may be seen from the cartoon in Fig \ref{fig:ws}.
In the case of $z_H<1/a$, the surface consists of two parts. The first part corresponds to the minimal surface in the conformal metric,  stretching from $z_M$ to $z_H$  is
\begin{equation}
\label{X1X}
     \Delta S_1=\frac{\sqrt{\lambda}}{2\pi}\!\!\int\limits_{-\mathcal T/2}^{\mathcal T/2}\!\!\!\dd\tau\int\limits_{z_M}^{z_H}\frac{2\dd z}{z^2}=\frac {\sqrt{\lambda}}{\pi} \,{\cal T}\,\bigg(\frac 1{z_M}-\frac 1{z_H}\bigg).
\end{equation}
The energy per particle, due to the presence of the hard wall for the first contribution is
\bea
\label{NOT}
\frac 12 E^{(1)}_{W}=\frac{\sqrt\lambda}{2\pi}\bigg(\frac 1{z_M}-\frac 1{z_H}\bigg).
\eea
It does not contribute to the entanglement, since it does not depend on the Unruh temperature. 

For the second contribution, we will evaluate the pertinent action. For that, we note that in Euclidean signature, (\ref{XHYPER}) is the spheroidal surface
\bea
\label{SWEAK}
\!-t^2\!+\!x^2\!+\!z^2\!=\!\frac{1}{a^2}\,\,\,\rightarrow\,\,\, \!t^2\!+\!x^2\!+\!z^2\!=\!\frac{1}{a^2}.
\eea
The contribution of the hard wall  reduces to the area of the disk at the radial position $z=z_H$
\bea
    \Delta S_2=&\frac{\sqrt\lambda}{2\pi z_H^2}\,\pi\left(\frac 1{a^2}-z_H^2\right)\,\,\,\rightarrow\,\,\, \frac{\sqrt\lambda}{2(az_H)^2},
\eea
with the rightmost term following from the weak field limit. The corresponding 
free energy is then  
\begin{equation}
\label{FREE}
    \beta_U F=\frac{\sqrt{\lambda}\,\beta_U^2}{8\pi^2\,z_H^2}.
\end{equation}
with $\beta_U=1/T_U$. The associated entanglement entropy is solely due to (\ref{FREE})
\begin{equation}
    S^{(2)}_\text{EE}=\beta_U^2\frac{\partial F}{\partial \beta_U}=\frac{\sqrt{\lambda}}{2(az_H)^2}.\label{eq:Confweak}
\end{equation}
which can be recast as
\bea
S^{(2)}_{EE}=\frac{\pi \sigma_T M^2}{E^2}.
\eea

\section{Holographic set-up: strong field}
\label{SEC_III}
The strong electric field analysis of the holographic Schwinger model was initially discussed in the conformal  AdS case in~\cite{Gorsky:2001up}, and revisited in~\cite{Semenoff:2011ng}. The confining 
and non-conformal case (Sakai-Sugimoto model) was analyzed in~\cite{Kim:2008zn}. In this section, we follow
~\cite{Semenoff:2011ng} and review briefly their construction and results in the conformal case, and then 
proceed to extend them to the confining case using walled AdS. 

\subsection{AdS$_5$}
The non-confining vacuum is unstable against $q\bar q$ pair creation for any finite electric field.
The creation probability  is exponentially suppressed by the action of the tunneling pair through 
the potential barrier as illustrated in Fig~\ref{fig:Veff} (bottom). In flat space and at weak coupling, this action follows from the worldline instanton, the cyclotron path traced by a charged quark, in a fixed external electric field, which acts much like a magnetic field in Euclidean signature. The result, is the famed Schwinger suppression factor.

At strong coupling, the holographic construction suggests that the action is that of a world-sheet instanton traced by the cyclotron path at the boundary, for infinitely massive quarks. For quarks with finite mass $M$, the cyclotron path is a Wilson loop on a D3 brane fixed at $z=z_M$ with $$M=\frac{\sqrt\lambda}{2\pi z_M}\,$$
as measured by the string hanging from D3 to the Poincare horizon. The string 
worldsheet in AdS with Euclidean signature  is described by the standard Nambu-Goto action~\cite{Semenoff:2011ng} 
\bea
    S_\text{NG}=&&\frac{\sqrt \lambda}{4\pi}\int\limits_{0}^1\dd\tau\,\int\limits_{\sigma_M}^\infty\dd\sigma\frac{1}{z^2}\left(\partial X_\mu\,\bar\partial X_\mu+\partial z\,\bar\partial z\right)\nonumber\\
    &&+i\oint A,\label{eq:classs}
\eea
where $(\partial,\bar\partial)=(\partial_\sigma\pm i \partial_\tau)$, with the additional Virasoro constraints. The last contribution has support only on the D3 boundary, and captures the effect of the Lorentz force with $F=dA$, on the source quarks,
\bea
\label{WLD}
x^M(\tau)=(R\,{\rm cos}(2\pi n\tau), R\,{\rm sin}(2\pi n\tau), 0,0, z_M)
\eea
tracing n-times a circular Wilson loop of classical cyclotron radius $$R\sim \frac 1a=\frac ME$$ for weak electric fields. For strong fields, the radius shrinks critically to zero (see below).
Note that the parameter shift $\tau\to 1/4-a\tau/2\pi$ in (\ref{WLD}) brings~\eqref{WLD} into~\eqref{BOUNDX}.

The variational equations following from (\ref{eq:classs}) and subject to the Virasoro constraints 
were detailed in~\cite{Semenoff:2011ng}, with the explicit worldsheet solution
\begin{widetext}
\bea
    X^M(\tau, \sigma)=\bigg(R\,{\rm cos}(2\pi n\tau)
    \frac{\cosh(2\pi\,n\,\sigma_M)}{\cosh(2\pi\,n\,\sigma)},
    R\,{\rm sin}(2\pi n\tau)
    \frac{\cosh(2\pi\,n\,\sigma_M)}{\cosh(2\pi\,n\,\sigma)},0,0,
    z_M\frac{\tanh(2\pi\,n\,\sigma)}{\tanh(2\pi\,n\,\sigma_M)}\bigg),
    \label{eq:classsol}
\eea
\end{widetext}
and the condition $\sinh(2\pi\,n\,\sigma_M)=z_M/R$. 
\begin{figure}
    \centering
    \includegraphics[width=0.6\linewidth]{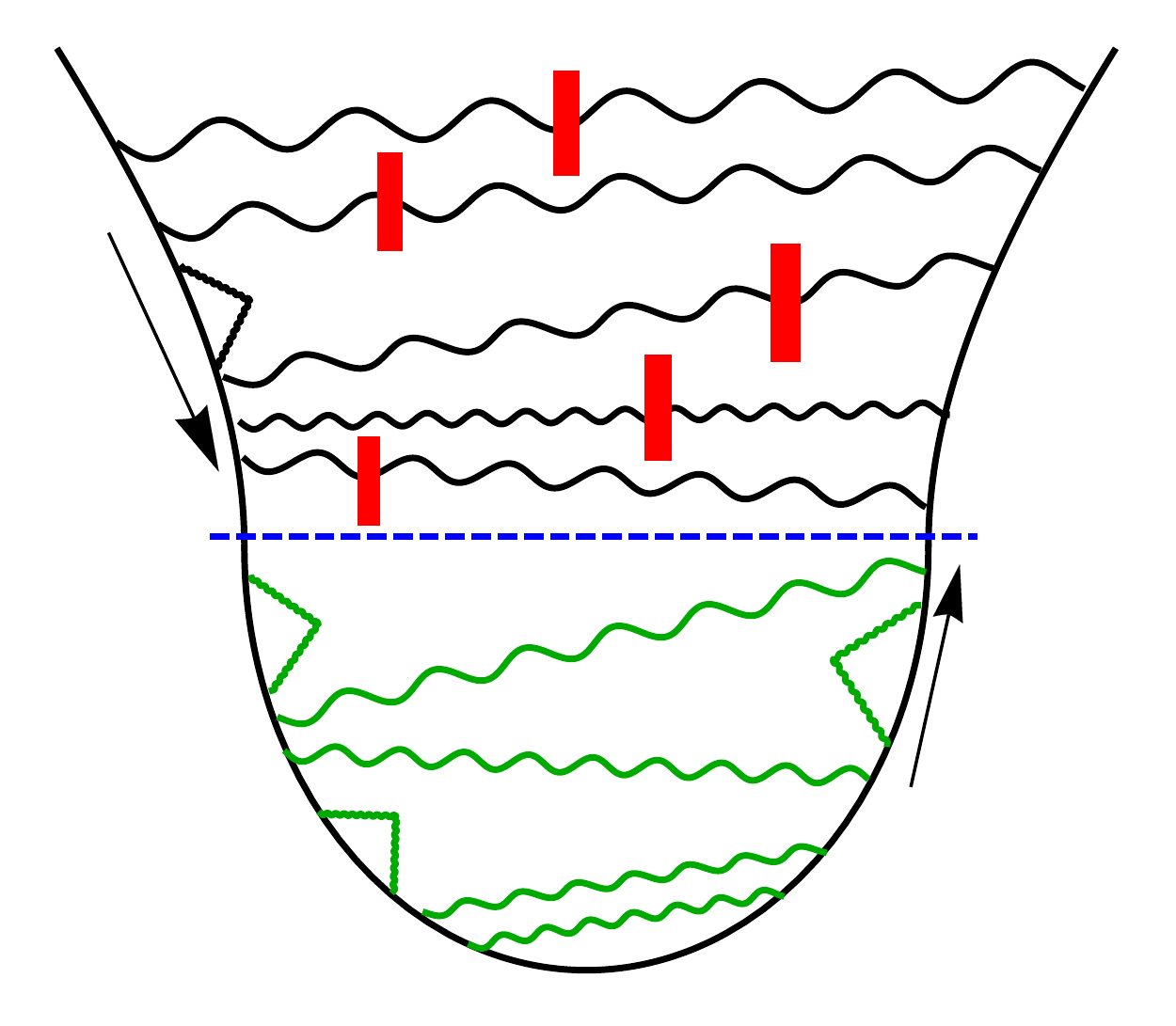}
    \caption{
    The trajectories of  a produced $q\bar q$ pair, with the tunneling part in Euclidean signature (lower half). The dashed blue line marks $t=0$ where the pair appears in Minkowski space. The virtual radiation is depicted in black and green,  and the (real) radiation leading to information loss is marked  in red, both in the large $N_c$  planar approximation. The red cut indicates that the exchanged gluon goes on-shell as radiation.}
    \label{fig:fey}
\end{figure}
The line-element corresponding to the worldsheet instanton (\ref{eq:classsol}) is
\bea
ds^2_W=\frac 1{{\rm sinh}^2(2\pi n\sigma)}(d\tau^2+d\sigma^2)
\label{XWS1}
\eea
Note that (\ref{XWS1}) develops a singularity  at $\sigma\rightarrow \infty$, 
\bea
ds^2_{W}\rightarrow  4{e^{-{4\pi n\sigma}}}(-d\tau^2+d\sigma^2)
\eea
after the analytical continuation to Minkowski signature 
$\tau\rightarrow i\tau$. This is the conformal line element of a black hole, with Rindler temperature $T_R=2\pi$ (after rescaling the affine parameters by 2).

Inserting (\ref{eq:classsol}) into (\ref{eq:classs}) yields  the on-shell action
\bea
\label{eq:onshell}
S_\text{on-shell}={n\sqrt\lambda}\bigg(\bigg(\frac {R^2}{z_M^2}+1\bigg)^{\frac 12}-1\bigg)-\frac{1}{2}(2\pi\,n)E R^2.\nonumber\\
\eea
The cyclotron radius $R$ is now fixed  by the extremum of~\eqref{eq:onshell}, which satisfies
\begin{equation}
    R^2+z_M^2=\frac 1{a^2}
    \label{eq:radius}
\end{equation}
The critical electric field is the electric field for which the cyclotron radius $R$ shrinks to zero, i.e.
$$z_Ma_c=z_M\frac {E_c}M=1\,.$$
In terms of the critical electric field $E_c$ (\ref{eq:onshell}), the on-shell action reads~\cite{Semenoff:2011ng}
\begin{equation}
    S_\text{on-shell}=\frac{\sqrt{\lambda}}{2}\left(\sqrt{\frac{E_c}{E}}-\sqrt{\frac{E}{E_c}}\right)^2,\label{eq:solon}
\end{equation}
This is consistent with Schwinger$^\prime$s result $S\rightarrow \frac{\sqrt{\lambda}\,E_c}{2E}$ in the weak field limit.  
In the strong field limit, the end-points of the string on the D3 boundary
accelerate substantially, as the cyclotron radius shrinks
\bea
a_e=\frac 1R
=\frac a{\sqrt{1-(a/a_c)^2}} .
\eea

If we recall that  the Unruh temperature is $T_U=\frac a{2\pi}$ 
for the pair as it tunnels out of the vacuum, then in Euclidean signature (\ref{eq:solon}) amounts to the 
free energy
\begin{equation}
    \frac {F}{T_U}
    =\frac{\sqrt{\lambda}}{2}\frac{T_c}{T_U}\left(1-\frac{T_U}{T_c}\right)^2
\end{equation}
The entanglement entropy follows from thermodynamics, with 
\begin{equation}
    S^T_\text{EE}=\beta_U^2\frac{\partial F}{\partial\beta_U}=\sqrt{\lambda}\left(1-\frac{T_U}{T_c}\right)\label{eq:EEstrongnoconf}
\end{equation}
and $\beta_U=1/T_U$.
Formula (\ref{eq:EEstrongnoconf}) 
generalizes  (\ref{XSEEX}) to the strong field regime. It is  the tunneling 
contribution to the pair creation process. Note that it reduces to the causal 
contribution (\ref{XSEEX}) (when no real radiation is emitted) in the weak field limit, when $T_U \to 0$.

On the basis of our strong field result \eqref{eq:EEstrongnoconf}, and the fact that the result $S_{EE} = \sqrt{\lambda}/3$ holds for any field strength, we can deduce the effect of real radiation on the entanglement entropy in the strong field regime.
To reconcile the result for the total entanglement entropy $S_{EE} = \sqrt{\lambda}/3$ with the result
\eqref{eq:EEstrongnoconf}, we have to assume that the effect of real radiation on the entanglement entropy in the strong field regime is given by 
\bea
S^{R}_\text{EE}=-\frac 23 \sqrt\lambda+\sqrt\lambda \frac{T_U}{T_c} .
\label{Xnocausal}
\eea
The strong field calculation described above does not apply for $T_U>T_c$, when the pair production is not exponentially suppressed. This is because 
the string worldsheet has no support for $z<z_M$
(the string calculation breaks down for strictly $R=0$).
 (\ref{eq:EEstrongnoconf}) plus (\ref{Xnocausal}) yield the net result 
\begin{widetext}
\bea
\label{NET}
S_{EE}=S^T_{EE}+S_{EE}^{R}=
\sqrt{\lambda}\left(1-\frac{T_U}{T_c}\right)+\sqrt\lambda\left(-\frac 23 
+\frac{T_U}{T_c}\right)=
\frac {\sqrt\lambda}3 ,
\eea
\end{widetext}
in agreement with (\ref{XSEE}), even for strong fields. At $T_U=T_c$ the tunneling barrier disappears.
For $T_U>T_c$, both the tunneling and the worldsheet black hole  disappears
as the horizon moves  below the D3 brane with $\frac 1a <z_M$. 
The ensuing pair creation process turns ``classical'', with vanishing entanglement entropy.

The result (\ref{NET}) can be interpreted as follows: The first bracket is the contribution of the virtual gluon interactions to the quantum entropy of a
receding pairs, as captured by the tunneling worldsheet instanton in bulk (depicted by the green exchanges in Fig.~\ref{fig:fey} below the blue-dashed  line). The second bracket
is the contribution to the entanglement entropy from the real emission of gluons by the pair  
(depicted by the black exchanges with red marks in Fig.~\ref{fig:fey}  above the blue-dashed  line).

\subsection{Walled AdS$_5$}
In walled AdS$_5$, the string worldsheet terminates on the wall at $z=z_H$, provided that the 
hard wall occurs prior to the effective horizon $z_M<z_H\leq \frac 1a$. 
Using (\ref{SWEAK}),
the contribution of the hard wall  is given by the area of the disk at the radial position $z=z_H$
\bea
    \Delta S_2=
    \frac{\sqrt\lambda}{2\pi z_H^2}\,\pi\left(\frac 1{a^2}-z_H^2\right)
\eea
The corresponding free energy is
\begin{equation}
    \beta_U F=\frac{\sqrt{\lambda}}{2}\left(\frac{\beta_U^2}{(2\pi\,z_H)^2}-1\right).
\end{equation}
from which the  entanglement entropy reduces to
\begin{equation}
    S^{(2)}_\text{EE}=\beta_U^2\frac{\partial F}{\partial \beta_U}=\frac{\sqrt{\lambda}}{2}\left(\frac{1}{(az_H)^2}+1\right)\label{eq:EEp2}
\end{equation}

To  evaluate  the contribution of the side of the string as it reaches the hard wall, we consider the action~\eqref{eq:classs} and the solution~\eqref{eq:classsol}. The string reaches the hard wall at $z=z_H$. In the world-sheet coordinates, the string is parametrized by a continuous function ranging from the starting point of the string at the boundary at $z=z_M$ $(\sigma=\sigma_0)$ to the turning point of the string at $z=z_\text{max}=\frac 1a$ $(\sigma=\infty)$. On the worldsheet this is related to 
$$z/z_M=\tanh(2\pi\, n\, \sigma)/\tanh(2\pi\, n\, \sigma_M)$$ 
or equivalently
$$2 \pi  n\sigma=\arctanh \left(\frac{z \tanh (2 \pi  n \sigma_M)}{z_M}\right)\,.$$ 
There are two different scenarios. If the cutoff is larger than the turning point of the string $z_H>z_\text{max}$, the hard wall does not affect the string, and we find the solution of the last section. If $z_H<z_\text{max}$, we cut off part of the string at 
$$\sigma_H=\arctanh \left(\frac{z_H \tanh (2 \pi  n \sigma_M)}{z_M}\right)/(2 \pi  n)\,.$$
Replacing the upper integration limit in~\eqref{eq:classs} by $\sigma=\sigma_H$, we have
\begin{widetext}
\bea
    S_\text{on-shell}=n\sqrt\lambda \bigg(\frac 1{az_M}-\frac 1{az_H}\bigg)-\frac 12 (2\pi n) ER^2
    =\frac{1}{2} \sqrt{\lambda } n \left(\frac{E}{E_c}+\frac{E_c}{E}-\frac{2 M }{E\,z_H}\right).
    \label{eq:onshellconf}
\eea
Since the cutoff is a hard wall, the radius is still given by~\eqref{eq:radius}.
For $z_H=z_\text{max}=\frac 1a$, the result reduces to~\eqref{eq:solon}. 
\end{widetext}

The free energy following from the on-shell action~\eqref{eq:onshellconf} for strong fields,  is
\begin{equation}
\!    F\!=\!\beta_U^{-1} S_\text{on-shell}\!=\!\frac{\sqrt{\lambda } n \left(\pi  \left(\beta_U ^2+\beta_c^2\right)-\beta_U ^2 \beta_c/z_H\right)}{2 \pi  \beta_U^2  \beta_c}.
\end{equation}
The contribution to the entanglement entropy is thus given by
\begin{equation}
    S_\text{EE}^{(1)}=\beta_U^2\frac{\partial F}{\partial \beta_U}=-n\sqrt{\lambda }\, \frac{\beta_c }{\beta_U }.\label{eq:EEp1}
\end{equation}
Adding the two contributions in~(\ref{eq:EEp1}) and~(\ref{eq:EEp2}), we find for a single winding with $n=1$
\begin{widetext}
\begin{equation}
   S_\text{EE}=S_\text{EE}^{(1)}+  S^{(2)}_\text{EE}=\begin{cases}
   \sqrt{\lambda}\left(\frac{M^2}{2(Ez_H)^2}+\frac12- \frac{E }{E_c}\right)\quad&\text{for: } 
   z_H\le M/E=\frac 1a\\ \frac 13\sqrt\lambda
   \quad&\text{for: } z_H>M/E=\frac 1a\end{cases}.\label{eq:eEEconf}
\end{equation}
\end{widetext}
with $z_H$ fixed by the string tension in (\ref{SIGMAT}), i.e.
\bea
\frac{\sqrt{\lambda}}2\frac{M^2}{(Ez_H)^2}= \frac {\sigma_T}{E}\frac{\pi M^2}E\equiv \frac{\sigma_T}{E}\,{\Delta}\rightarrow \frac 14 \Delta\,.
\label{D8}
\eea
The rightmost result follows from the estimate for the invariant electric field, as a source for the string tension $E\sim 4\sigma_T$~\cite{Casher:1978wy}. Here $\Delta$ is the Schwinger tunneling action for pair creation, with probability
$e^{-\Delta}$.
For a comparison, we note the corresponding Shannon (information) entropy 
for a single tunneling process
\bea
\label{SHAN}
S_{I}=\Delta\,e^{-\Delta}-(1-e^{-\Delta}){\rm ln}\,(1-e^{-\Delta})
\rightarrow \Delta 
\eea
which is comparable to (\ref{D8}) in the weak field limit (rightmost result).

\begin{figure}
    \centering
    \includegraphics[width=0.8\linewidth]{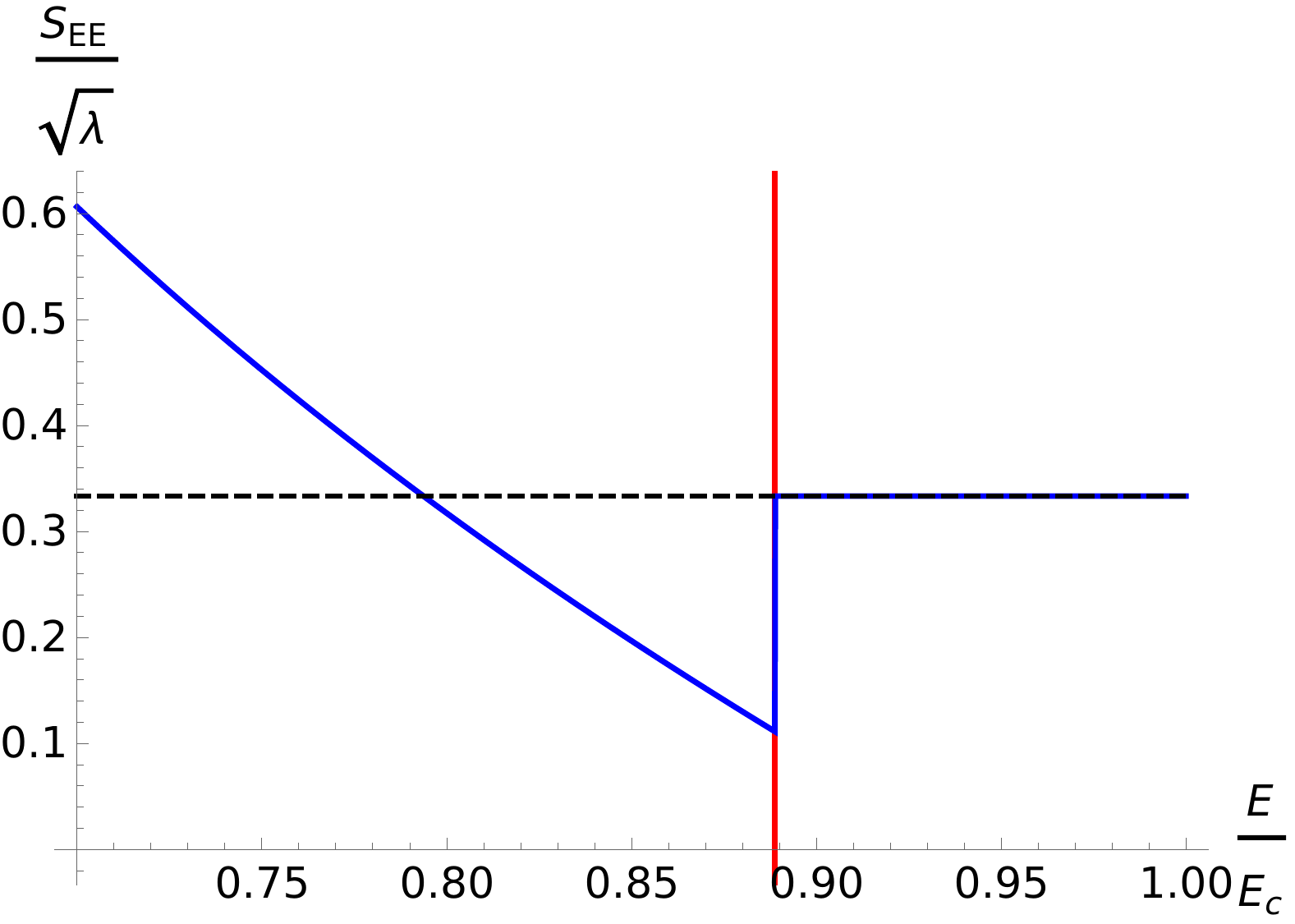}
    \caption{Entanglement entropy (\ref{eq:eEEconf}) versus $E/E_c$, in the confining background (blue-solid curve) and the deconfined background (black-dashed curve), for a hard wall at $z_H=0.89/E_c$ in units of $M=1$. For $E>1/z_H$, the entanglement entropies merge. The first order jump reflects on the transition from a confining to a black-hole geometry. }
    \label{fig:1BH}
\end{figure}

\begin{figure}
    \centering
    \includegraphics[width=0.8\linewidth]{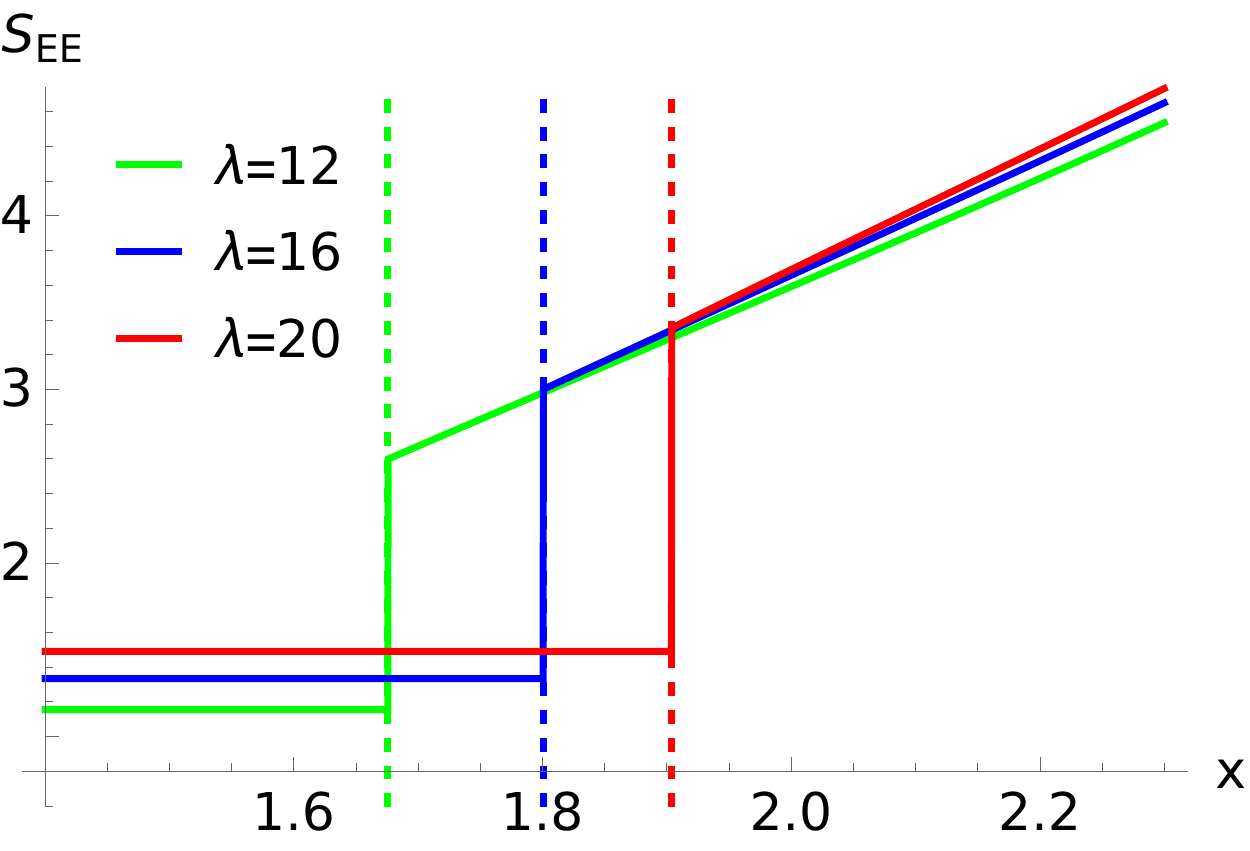}
    \caption{Entanglement entropy~\eqref{eq:eEEconfX} as function of the mass ratio $x=M/m_\rho$ for different $^\prime$t Hooft coupling $\lambda$. The dashed lines indicate $x=\frac{4\lambda^{\frac 14}}{\pi \sqrt 2}$.}
    \label{fig:4}
\end{figure}

In the limiting case of $z_H=1/a$, the entanglement entropy in~(\ref{eq:eEEconf}) correctly reproduces the entanglement entropy computed in~(\ref{eq:EEstrongnoconf}). Furthermore, we correctly reproduce the weak field limit in the deconfined case given by~(\ref{XSEEX}),  as well as the weak field confined case in~(\ref{eq:Confweak}). In Fig.~\ref{fig:1BH} we show the behavior of (\ref{eq:eEEconf}) 
versus $E/E_c$. The entanglement is larger in the confining case for weak fields with no radiation loss present, and decreases  linearly as the electric field increases following the depletion of the tunneling process.  The first order jump occurs when the geometry flips from confining to black-hole. The location of the jump varies with $M$.
\\
\\
{\bf General case:}
\\
In general for weak electric fields (low Unruh temperature), the entanglement entropy appears to diverge. However, there is a lower bound on the electric field determined by the  string tension $\sigma_T$ 
\begin{equation}
    \frac{T_\text{min}}{T_c}=\frac{E_\text{min}}{E_c}>\frac{\sigma_T}{E_c}=\frac{\sqrt{\lambda}}{4\pi^2\alpha' M^2}.
\end{equation}
To make a contact with QCD phenomenology, we use
$\alpha'm_\rho^2=\frac 12 $ (where $m_\rho$ is the $\rho$ meson mass), and find a light constituent quark mass $M$, with $M/m_\rho\sim \frac 12 $. If we set $\lambda=g_Y^2N_c\approx 12$, then the lower bound is
\begin{equation}
    \frac{T_\text{min}}{T_c}=\frac{E_\text{min}}{E_c}
    >\frac{2\sqrt{\lambda}}{\pi^2}\approx\frac{4\sqrt{3}}{\pi^2}\approx0.7.
\end{equation}
In Fig.~\ref{fig:1BH}, we display the final result (\ref{eq:eEEconf}) for the entanglement entropy in the confining geometry (blue-solid curve), and the geometry without hard wall (dashed-black curve) versus $E/E_c$ in units of $M=1$. The hard wall is placed at a fixed  holographic distance $z_H=0.89/E_c$. 
The entanglement entropy is dominated by confinement for $E/E_c<0.89$, and merges with the entanglement entropy in conformal AdS, as the  ER bridge overrides the hard wall in bulk (red line).
\\
\\
{\bf Special case:}
\\
For the QCD string, the invariant electric field is about $E\sim 4\sigma_T$~\cite{Casher:1978wy}. If we also fix the string tension to reproduce the rho meson trajectory $\sigma_T=\frac 12 m_\rho^2$, with a varying mass ratio $x=M/m_\rho$, and a fixed $^\prime$t Hooft coupling $\lambda$, then (\ref{eq:eEEconf}) reduces to
\begin{widetext}
\begin{equation}
   S_\text{EE}(x,\lambda)=\begin{cases}
   \frac{\pi^2}{16} x^2+\frac{\sqrt\lambda}2-\frac{2\lambda}{\pi^2 x^2}
   \quad&\text{for: } x\geq \frac{4\lambda^{\frac 14}}{\pi \sqrt 2}\\\
   \frac 13\sqrt\lambda
   \quad&\text{for: } x< \frac{4\lambda^{\frac 14}}{\pi \sqrt 2}
  \end{cases}.\label{eq:eEEconfX}
\end{equation}
\end{widetext}
In fig~\ref{fig:4}, we illustrate the dependence of the entanglement entropy~\eqref{eq:eEEconfX} on the mass ratio $x$ for different values of the $^\prime$t Hooft coupling $\lambda$. For smaller $\lambda$, we can reach smaller mass ratios and the de-confinement transition is moving to smaller values of the mass ratio $x$.

\section{Entropy in the Lund model}
\label{LUNDX}
A simple but phenomenologically successful description  of a hadronization process in QCD,
is the Lund model~\cite{Andersson:1983ia}, which makes use of the QCD string. The model  describes the break-up of the string into hadrons. For instance, as an initial  $q\bar q$ pair is produced by a hard jet, the connecting string stretches as the pair recedes away from each other. Eventually,
several $q\bar q$ pair  pop up by tunneling, and the stretched  string breaks into  shorter strings, identified as hadrons (mostly mesons).

A key feature of the multiple break-ups is the so-called area law, at the origin of the probability  distribution in the Lund model. The area law is a product of Schwinger tunneling like  probabilities,
as we have described in the confining case above. 
Schematically, the normalized probability for the string breaking to n-identical hadrons (say pions)   in 1+1 dimensions, is of the form
\bea
\label{PN1}
p_n=(e^{A}-1)e^{-nA}\approx 
(e^{\frac 1{\bar n}}-1)e^{-\frac n{\bar n}}
\eea
Here  $$A\sim \frac {M^2}{\sigma_T}\sim 
{\Delta}$$ is a typical fragment area of the string  worldsheet (in units of the string length squared),
with $M$ the light quark constituent mass at the string end points. Note that the mean multiplicity $\bar n$ and $A$, are tied by 
\bea
A=-{\rm ln}\bigg(1-\frac 1{\bar n}\bigg)\approx \frac 1{\bar n}\,,
\eea
which shows that (\ref{PN1}) obeys Koba-Nielsen-Olesen (KNO) scaling for large multiplicities 
\bea
\bar n p_n\approx e^{-\frac n{\bar n}}
\eea
Remarkably, this is a distribution of a thermal oscillator of temperature $T=\bar n/\omega$, with $\omega$ in units of the string length.
The corresponding Shannon (information) entropy is
\bea
\label{PN2}
S_{EE}=-\sum_n p_n\,{\rm ln}\,p_n\approx {\rm ln}\,\bar n 
\eea
which we identify as the entanglement entropy of the jet fragmentation. It
is a measure of the Schwinger tunneling probability  (as weighted by the mean area of the string fragment). Note that the multiple breaking of the string in the Lund model
reduces the entanglement entropy compared to the single string breaking  \eqref{D8}, from $\frac 14 \Delta$ to ${\rm ln}\frac 1\Delta$. The reduction underlines the quantum  decoherence caused by the many random breaking on the string worldsheet, spanned by the initial receding 
pair from the jet. It is analogous to the reduction of quantum entanglement by radiation discussed above.

The result (\ref{PN2}) is in agreement with the arguments presented in~\cite{Liu:2022bru,Liu:2023eve} for jet fragmentation at large invariant mass. This 
result was initially noted for $pp$ and $ep$ at small Bjorken $x$~\cite{Stoffers:2012mn,Kharzeev:2017qzs}. The KNO-based  arguments confirm the identification of
(\ref{PN2}) with the entanglement entropy for QCD, as computed from QCD evolution equations \cite{Kharzeev:2017qzs}. The fact that (\ref{PN2}) is also recovered for the schematic form of the Lund model discussed here, underlines the duality of the partonic and string descriptions of QCD. It also reflects on the Markovian character of both fragmentation processes.

\section{Conclusions}
\label{SEC_IV}

We have revisited the holographic Schwinger pair creation process in the simplest confining AdS geometry, a slab of AdS$_5$ with a wall at $z=z_H$. When $z_H$ is removed to infinity, confinement disappears and  the pair creation process in curved AdS initially discussed in~\cite{Semenoff:2011ng} is recovered for both weak and strong electric fields. As the pair recedes, a holographic string develops with an ER bridge forming in the bulk, encoding geometrically their quantum entanglement~\cite{Jensen:2013ora}. 

In a dynamical pair creation process, the quantum entanglement entropy is composed of a contribution from the string above the ER bridge, minus the contribution from the string 
as it enters the ER bridge~\cite{Lewkowycz:2013laa,Jensen:2014bpa}. The latter is dual to the real radiation in the Minkowski space on the boundary. Part of the contribution above the ER
bridge can be traced back to the tunneling process. It decreases linearly the stronger the electric field, and vanishes when the field reaches its critical value $E_c$, for which the potential barrier disappears. The process of pair creation becomes "classical", with no tunneling penalty.

In the presence of confinement, the pair creation process is altered. The creation 
process occurs  only for electric fields in the range $\sigma_T<E<E_c$. 
For sufficiently weak electric fields in comparison to the string tension $\sigma_T$, confinement 
inhibits  the pair creation process, as virtual pairs strongly bind in the vacuum. 

For moderately  weak fields in the range $\sigma_T<E\ll E_c$,  the quantum entanglement is stronger, since the confining scale overrides the ER bridge. The increase in the quantum entanglement of the produced pair
is caused by their binding through the string through the hard wall in bulk. For moderately strong electric fields in the range $\sigma_T\ll E< E_c$, the ER bridge overrides the hard wall, with the  quantum entanglement reducing to that of the conformal AdS space, and fixed by the causal and non-causal contributions delineated by the ER bridge.

Using the Lund model, as a prototype for string breaking in 1+1 dimensions, we have suggested that the Schwinger tunneling probability is  directly related to the entanglement entropy, through the KNO scaling.

The effects of strong coupling on quantum entanglement between the produced particles in time-dependent electric backgrounds is also of great interest. We will report on the results of the corresponding study in a forthcoming publication.

In recent years, the advent of intense lasers and energetic particle beams has led to substantial progress in the study of QED in intense background fields (see~\cite{Dunne:2004nc,Fedotov:2022ely} for reviews). One important non-perturbative phenomenon in this regime is the Schwinger effect. The effects of quantum entanglement in this process have to be explored, and we hope that our work is a step in that direction.
\newline\newline
{\noindent\bf Acknowledgments}

\noindent 
We thank Adrien Florio for useful discussions.
This work was supported by the Office of Science, Office of Nuclear Physics, U.S. Department of Energy under Contract No. DE-FG88ER41450 and No. DE-FG-88ER40388, and by the U.S. Department of Energy, Office of Science, National Quantum Information Science Research Centers, Co-design Center for Quantum Advantage (C2QA) under Contract No.DE-SC0012704 (DK).

\bibliography{references}
\end{document}